%% file: DTPK1_1.tex
\documentclass[12pt]{article}

\usepackage[dvips]{epsfig}
\usepackage{amsfonts}
\usepackage{amssymb}
\usepackage{amscd}
\usepackage{amstext}
\usepackage{amsmath}
\usepackage{pstricks}
\usepackage{rotating}
\usepackage{xy}
\xyoption{all}

\newtheorem{example}{Example(s)}[section]

\newtheorem{theorem}[example]{Theorem}

\newtheorem{definition}[example]{Definition}
\newtheorem{proposition}[example]{Proposition}

\newtheorem{lemma}[example]{Lemma}

\newtheorem{remarks}[example]{Remarks}
\newtheorem{note}[example]{Note}

\def\Proof{\medskip\noindent {\it Proof --- \ }}
\def\cqfd{\hfill $\Box$ \bigskip}

\def\hs{\hbox to 3mm{}}
\def\hhs{\hbox to 5cm{}}
\def\ss{\smallskip}

\def\bs{\bigskip}

\def\shuff#1#2{\mathbin{
      \hbox{\vbox{
        \hbox{\vrule
              \hskip#2
              \vrule height#1 width 0pt
               }%
        \hrule}%
             \vbox{
        \hbox{\vrule
              \hskip#2
              \vrule height#1 width 0pt
               \vrule }%
        \hrule}%
}}}
\def\shuffle{{\mathchoice{\shuff{7pt}{3.5pt}}%
                        {\shuff{6pt}{3pt}}%
                        {\shuff{4pt}{2pt}}%
                        {\shuff{3pt}{1.5pt}}}}

\def\adots{\mathinner{\mkern2mu\raise1pt\hbox{.}
\mkern3mu\raise4pt\hbox{.}\mkern1mu\raise7pt\hbox{.}}}

\def\pointir{\unskip . --- \ignorespaces}

\def\up#1{\raise 1ex\hbox{\footnotesize#1}}

\def\mref#1{(\ref{#1})}
\def\bu{\bullet}

\def\ra{\rightarrow}
\def\da{\downarrow}

\def\ua{\uparrow}

\def\A{\mathcal{A}}
\def\D{\mathcal{D}}

\def\N{{\mathbb N}}
\def\C{{\mathbb C}}

\def\X{{\mathbb X}}
\def\al{\alpha}
\def\be{\beta}
\def\ga{\gamma}
\def\ep{\epsilon}
\def\om{\omega}

\def\L{\mathbb{L}}
\def\V{\mathbb{V}}

\def\scal#1#2{\langle #1 | #2 \rangle}

\def\ncp#1#2{#1\langle #2\rangle}

\def\Pol{\mathrm{Pol}}

\def\diag{\mathbf{diag}}
\def\ldiag{\mathbf{ldiag}}

\def\LDIAG{\mathbf{LDIAG}}
\def\DIAG{\mathbf{DIAG}}

\def\MQS{\mathbf{MQSym}}

\begingroup
\def\2#1{\ifnum#1<10 0\fi\the#1}
\xdef\isodayandtime{
{\2\day-\2\month-\the\year\space\2{\count0}:%
\2{\count2}}}
\endgroup

\title{\Large\bf Combinatorial Deformations of Algebras:\\ 
Twisting and Perturbations}

\author{
{\sc G. H. E. Duchamp, C. Tollu,}   
\rm\thanks{LIPN - UMR 7030
CNRS - Universit\'e Paris 13
F-93430 Villetaneuse, France},\\
{\sc K. A. Penson}
\rm\thanks{Laboratoire de Physique Th\'eorique de la Mati\`{e}re Condens\'{e}e
Universit\'e Pierre et Marie Curie, CNRS UMR 7600
Tour 24 - 2i\`eme \'etage, 4 place Jussieu, F 75252 Paris cedex 05} and
{\sc G. Koshevoy}
\rm\thanks{Central Institute of Economics and Mathematics (CEMI)
Russian Academy of Sciences
},\\
}
\date{}

\begin{document}

\maketitle

\tableofcontents

\begin{abstract}
The framework used to prove the multiplicative deformation
of the algebra of Feynman-Bender diagrams is a \textit{twisted shifted
dual law} (in fact, twisted twice). We give here a clear interpretation of
its two parameters. The crossing parameter is a deformation of
the tensor structure whereas the superposition parameter is a
perturbation of the shuffle coproduct which, 
in turn, can be interpreted as the diagonal restriction of a 
superproduct. Here, we systematically detail these constructions.
\end{abstract}

\section{Introduction}
In \cite{BBM}, Bender, Brody, and Meister introduced a special field theory, then called ``Quantum Field Theory of Partitions''. This theory is based on a bilinear product formula which reads

\begin{eqnarray}
\mathcal{H}(F,G)=\left.F\left(z\frac{d}{dx}\right)G(x)\right|_{x=0}.
\end{eqnarray}

If one develops this formula in the case when $F$ and $G$ are free exponentials, one obtains a summation over all the (finite) bipartite \footnote{The (bi)-partition of the vertices is understood ordered. In this case, the term {\it bicoloured} can also be found in the literature.} graphs with multiple edges and no isolated point \cite{GoF12} (the set of these diagrams will be called $\diag$), a data structure which is equivalent to classes of packed matrices \cite{DHT} under permutations of rows and columns.\\
So, one has a Feynman-type expansion of the product formula 
\begin{eqnarray}
\mathcal{H}\left(\exp(\sum_{n=1}^\infty L_n\frac{z^n}{n!}),\exp(\sum_{n=1}^\infty V_n\frac{z^n}{n!})\right)=\sum_{n\geq 0} \frac{z^n}{n!} \sum_{d\in \diag\atop
|d|=n} mult(d)\L^{\al(d)}\V^{\be(d)}
\end{eqnarray}
where $mult(d)$ is the number of pairs $(P_1,P_2)$ of (ordered) set partitions of $\{1,\ldots,n\}$ which correspond to a diagram $d$, $|d|$ the number of edges in $d$ and 
\begin{equation}
	\L^{\al(d)}=L_1^{\al_1}L_2^{\al_2}\cdots \ ;\ \V^{\be(d)}=V_1^{\be_1}V_2^{\be_2}\cdots 
\end{equation}
is the multiindex notation for the monomials in $\L\cup \V$ where $\al_i=\al_i(d)$ (resp. $\be_j=\be_j(d)$) is the number of white (resp. black) spots of degree $i$ (resp. $j$) in $d$.\\
The set $\diag$ endowed with disjoint receive the structure of a monoid such that the arrow $d\mapsto \L^{\al(d)}\V^{\be(d)}$ is a morphism (of monoids) and then, by linear extension, one deduces a morphism of algebras 
\begin{equation}\label{multiplier_map}
	\C[\diag]\ra \Pol(\C;\L\cup \V)\ .
\end{equation}
where $\Pol(\C;\L\cup \V)$ is the Hopf algebra of (commutative) polynomials with complex coefficients generated by the alphabet $\L\cup \V$.
For at least three models of Physics, one can specialize $\L$ so that the canonical Hopf algebra structure of $\Pol(\C;\L\cup \V)$ can be lifted, through \mref{multiplier_map}. The resulting Hopf algebra (based on $\C[\diag]$) has been denoted $\DIAG$. To our great surprise, this Hopf algebra structure could be lifted at the (noncommutative) level of the objects themselves instead of classes, resulting in the construction of a Hopf algebra on (linear combinations of) ``labelled diagrams'' (the monoid $\ldiag$, see \cite{GoF12}). As these ``labelled diagrams'' are in one-to-one correspondence with the packed matrices of $\MQS$, we get on the vector space $\C[\ldiag]$ two (combinatorially natural) structures of algebra (and co-algebra) and one could raise the question of the existence of a continuous deformation between the two. The answer is positive and can be performed through a three-parameter (two formal, or continuous and one boolean) Hopf deformation\footnote{This algebra deformation has received recently another realisation in terms of bi-words \cite{SFCA07}.} of $\LDIAG$ called $\LDIAG(q_c,q_s,q_t)$ \cite{GoF12} such that 

\begin{equation} \LDIAG(0,0,0)\simeq \LDIAG\ ;\ \LDIAG(1,1,1)\simeq \MQS\ .
\end{equation}

The r\^ole of the two parameters $q_c,q_s$ (algebra parameters, whereas $q_t$ is a coalgebra parameter) was discovered just counting crossings and superpositions in the twisted labelled diagrams (see \cite{GoF12} for details). This simple statistics (counting crossings and superpositions) yields an associative product on the diagrams. The first proof given for the associativity was mainly computational and it was a surprise that even the associativity held. 
This raised the need to understand this phenomenon in a deeper way and the question whether the two parameters ($q_c$ and $q_s$) would be of different nature. The aim of this paper is to answer this question and give a conceptual proof of associativity by developing four building blocks which are general and separately easy to test: addition of a group-like element to a co-associative coalgebra, shifting lemma, codiagonal deformation of a semigroup and extension of a colour factor to words.\\
The essential ingredient in the two last operations is what has become nowadays a useful tool, the coloured product of algebras, for which we give some new results.

\ss
{\sc Acknowledgements} : The authors are pleased to acknowledge the hospitality of institutions in Moscow and New York. Special thanks are due to Catherine Borgen for having created a fertile atmosphere in Exeter (UK) where the first and last parts of this manuscript were prepared. We take advantage of these lines to acknowledge support from the French Ministry of Science and Higher Education under Grant ANR PhysComb. We are also grateful to Jim Stasheff for having raised the question of the different natures of the parameters $q_c$ and $q_s$.

\section{The deformed algebra $\LDIAG(q_c,q_s)$}
\subsection{Review of the construction of the algebra}

The complete story of the algebra of Feynman-Bender diagrams which arose in Combinatorial Physics in (2005) can be found in \cite{GoF12} and a fragment of it, as well as a realization with an alternative data structure, in \cite{SFCA07}. 

\ss
Recall that (classical) shuffle products (of words) can be expressed in two ways\\
a) recursion\\
b) summation on (and by means of) some permutations.\\

Here, we will trace back the construction of the deformed product between two diagrams, starting from an analog of (b) (using however the symmetric semigroup instead of the symmetric group, see below) and going gradually to (a) following in that the first description of the deformed case which was graphical (and was discovered as such \cite{GoF12}).\\
The diagrams on which the product has to be performed are plane bipartite graphs (vertices being called black and white spots) with multiple ordered edges ; they look as follows.  

\bs
\input{DTPK_fig1}

\bs


One can define more formally this data structure using the equivalent notion of a weight function. Here, it is a function $\om : \N^+\times\N^+\ra \N$ (as in \cite{SFCA07}) with support 
\begin{equation}
supp(\om)=\{(i,j)\in \N^+\times\N^+\ |\ w(i,j)\not=0\}
\end{equation}
having projections of the form $pr_1(supp(\om))=[1\ldots p];\ pr_2(supp(\om))=[1\ldots q]$ for some $p,q\in \N^+$. This last prescription can be rephrased without $pr_i$ remarking that $p$ (resp. $q$) is the last $i$ such that $(\exists j\in \N^+)(\om(i,j)\not=0)$ (resp. $q$ is the last $j$ such that\\ $(\exists i\in \N^+)(\om(i,j)\not=0)$). In this way our graphs are in one-to-one correspondence with such weight functions. 

\bs
\input{DTPK_fig2}

\bs
We are now in the position of describing the (deformed) product of our diagrams by means of the symmetric semigroup (whereas the symmetric group would only provide crossings as it occurs with the shuffle product). 

\ss
The symmetric semigroup on a finite set $F$ (denoted here $SSG_F$) is the set of endofunctions $F\ra F$. In order to preserve the requirement that black spots kept on being labelled from $1$ to some integer, we have to ask that the mapping acting on the diagram $d$ with $n$ black spots had its image of the type $[1\ldots m]$ for some $m\leq n$. The result noted $d.f$ has $m$ black spots such that the black spot of (former) label ``$i$'' bears the new label $f(i)$.\\
If we consider any onto mapping $[1\ldots p]\ra [1\ldots r]$, the diagram $d.f=d'$ has the following weight function $\om'$
\begin{equation}
	\om'(i,k)=\sum_{f(j)=i} \om(j,k)\ .
\end{equation}
which can be easily checked to be admissible in our context.

Before giving the expression of the deformed product, we must define local partial degrees. 
For a black spot with label ``$l$'', we denote by $bks(d,l)$ its degree (number of adjacent edges).
Then, for $d_1$ (resp. $d_2$) with $p$ (resp. $q$) black spots, the product reads
\begin{equation}
	[d_1|d_2]_{L(q_c,q_s)}=\sum_{f\in Shs(p,q)} \Big(\prod_{i<j\atop f(i)>f(j)} q_c^{bks(d,i).bks(d,j)}\Big)\Big(\prod_{i<j\atop f(i)=f(j)} q_s^{bks(d,i).bks(d,j)}\Big) [d_1|d_2]_{L}.f
\end{equation}
where $Shs(p,q)$ is the set of mappings $f\in SSG_{[1\ldots p+q]}$ with image an interval of type $[1\ldots m]$ (with $\mbox{max}\{p,q\}\leq m\leq p+q$), 
and such that 
\begin{equation}
	f(1)<f(2)<\cdots <f(p)\ ;\ f(p+1)<f(p+2)<\cdots <f(p+q)\ .
\end{equation}
This condition, similar to that of the shuffle product, guarantees that the black spots of the diagrams are kept in order during the process of shuffling with superposition (hence the name $Shs$).

\subsection{Coding and the recursive definition}\label{code_to_rec}

The graphical and symmetric-semigroup-indexed description of the deformed product neither give immediately a recursive definition nor an explanation of ``why'' the product is associative. We will, on our way to understand this (as well as the different natures of its parameters), proceed in three steps:
\begin{itemize}
	\item coding the diagrams by words of monomials
	\item presenting the product as a shifted law
	\item give a recursive definition of the (non-shifted) law.
\end{itemize}

The code used here relies on monomials over a commutative alphabet of variables\\
 $\X=\{x_i\}_{i\geq 1}$. As in \cite{GoF12}, we let $\mathfrak{MON}(\X)$ denote the monoid of monomials $\{\X^\al\}_{\al\in \N^{(\X)}}$ (indeed, the free commutative monoid over $\X$) and $\mathfrak{MON}^+(X)=\{\X^\al\}_{\al\in \N^{(\X)}-\{0\}}$ the semigroup of its non-unit elements (the free commutative semigroup over $\X$).\\
Note that each weight function $\om\in \N^{(\N^+\times \N^+)}$ yields an equivalent ``word of monomials\footnote{The low point $.$ here is used to emphasize concatenation which is alsewhere denoted by simple juxtaposition of letters.}'' $W(\om)=w_1.w_2.\cdots .w_p$	such that 
\begin{equation}
	W(\om)[i]=w_i=\prod_{j=1}^\infty x_j^{\om(i,j)}\ .
\end{equation}
The correspondence $code\ :\ \N^{(\N^+\times \N^+)}\ra (\mathfrak{MON}^+(X))^*$ is one-to-one and provides at once a way to code each labelled diagram through its weight function as a word of monomials. Conversely a word $W\in (\mathfrak{MON}^+(X))^*$ is the code of a diagram (i. e. the image by $code$ of the weight function of a diagram) iff 
\begin{equation}\label{criterium_code_of_a_diagram}
	indexes(Alph(W))=[1\ldots m]
\end{equation}
(where {\footnotesize$indexes(Alph(W))$} is the set of $i\in \N^+$ such that an $x_i$ is involved in $W$).
Due to the special indexation of its alphabet, the monoid $(\mathfrak{MON}^+(X))^*$ comes equipped with a set of endomorphisms, the translations $T_n$ defined on the variables by $T_n(x_i)=x_{i+n}$ and extended to $\mathfrak{MON}^+(X)$, to $(\mathfrak{MON}^+(X))^*$ and then to $\ncp{K}{\mathfrak{MON}^+(X)}$.\\ 
Note that the code of a concatenation reads
\begin{equation}
	code([d_1|d_2]_L)=code(d_1).T_{max(indexes(Alph(code(d_1))))}(code(d_2))\ .
\end{equation}
Therefore, the function ``$code$'' being below extended by linearity, the reader may check easily that one can compute recursively the deformed product on the codes by 
\begin{equation}
	code([d_1|d_2]_L)=code(d_1)\ua T_{max(indexes(Alph(code(d_1))))}(code(d_2))
\end{equation}
where the bilinear product $\ua$ is recursively defined on the words as follows

\begin{eqnarray}\label{first_recursion}
\left\{ \
\begin{array}{rcl}
1_{(\mathfrak{MON}^+(X))^*}\ua w & = & w\ua 1_{(\mathfrak{MON}^+(X))^*}=w\\
au\ua bv &=& a(u\ua bv)+ q_c^{|au||b|}b(au\ua v) + q_c^{|u||b|}q_s^{|a||b|} (a\cdot b)(u\ua v)
\end{array}
\right.
\end{eqnarray}
where $a\cdot b$ (medium $\cdot$ dot) denotes the (monomial, commutative) product of $a$ and $b$ within $\mathfrak{MON}^+(X)$.

It is this last recursion that we will decompose and analyse below in order to get a better understanding of the parameters.\\
The associativity of the product \mref{first_recursion} is a consequence of the following proposition.

\begin{proposition}\label{deformed_law} (Prop. 5.1 in \cite{GoF12}) Let $(S,.)$ be a semigroup graded by a degree function $|\ |_d:\ S\ra \N$ (i. e. a morphism to $(\N,+)$)
and $S^*$ the set of lists (denoted by words $a_1a_2 \cdots a_k$) with letters in $S$ (including the empty list $1_{S^*}$).\\

Let $q_c,q_s\in K$ be two elements in a (commutative) ring $K$. We define on\\
$\ncp{K}{S}=K[S^*]$ a new product $\ua$ by
\begin{eqnarray}\label{deformed_infiltr}
w\ua 1_{S^*}&=&1_{S^*}\ua w=w\cr 
\hspace{-15mm} 
au\ua bv&=& a(u\ua bv)+ q_c^{|au|_d|b|_d} b(au\ua v)+ 
q_c^{|u|_d|b|_d}q_s^{|a|_d|b|_d} (a . b)(u\ua v)
\end{eqnarray}
where the weights are extended additively to
lists (words) by
$$
\Big|a_1a_2\cdots a_k\Big|_d=\sum_{i=1}^k |a_i|_d\ .
$$
Then the new product $\ua$ is graded, associative with $1_{S^*}$ as its unit.
\end{proposition}

The questions that have arisen in the introduction can be now reformulated as follows.

Q1) Are $q_c$ and $q_s$ of the same nature ?

Q2) If no, can the associativity be explained, step by step, 
by constructions which will show their different natures ?\\

Here, by{\it ``nature''}, is understood that, although at the level of statistics $q_c$ and $q_s$ seem to play a similar r\^ole, they could be distinguished by general algebra. Indeed we attempt in the sequel to show that $q_c$ is of geometric nature (deformation at the level of the tensor structure) whereas $q_s$ is a perturbative nature (perturbation of the Lie coproduct).\\
With this end in view, we need to recall a now classical tool, the coloured product of two algebras.  

\section{Colour factors and products}

Colour factors were introduced\footnote{In fact, some of them ({\it ``Facteurs de commutation''}, with values in $\{-1,1\}$ and an [anti]symmetry condition) are already considered in the edition of 1970 of \cite{B_Alg_I_III}. See the paragraph 10 {\it D\'erivations} of Ch III.} by \cite{Ree} and the theory was developped or used in \cite{ncsf3,D15,MZ,JZ}.

Let $\displaystyle{\mathcal A={\oplus}_{\alpha\in \mathcal
D}\mathcal A_\alpha}$ and $\displaystyle{\mathcal
B=\oplus_{\beta\in \mathcal D}\mathcal B_\beta}$ be two $\mathcal
D$-graded associative $K$-algebras\footnote{Not necessarily with unit.} ($\mathcal D$ is a commutative semigroup whose law is denoted additively). 
Readers that are not familiar with graded algebras can think of $\mathcal D=\N^{(X)}$, the free commutative monoid over $X$ and $\displaystyle{\mathcal A_\alpha=K[X]_\alpha}$, the space of homogeneous polynomials of multidegree $\alpha$.

Given a mapping $\chi :\mathcal D\times\mathcal D\longrightarrow K$, we define a product of algebra on $\mathcal{A}\otimes \mathcal{B}$ by

\begin{eqnarray}
  (x_1\otimes y_1)(x_2\otimes y_2) &=& \chi(\beta_1,\alpha_2)(x_1x_2\otimes
  y_1y_2)\label{prod_tens_coloré}
\end{eqnarray}

for $\displaystyle{(x_i)\in\mathcal A_{\al_i}}$ and $\displaystyle{(y_i)\in\mathcal B_{\be_i}}$ ($i=1,2$).\\
Equating the computations of $\left((x_1\otimes y_1)(x_2\otimes y_2)\right)(x_3\otimes y_3)$ and 
$(x_1\otimes y_1)\left((x_2\otimes y_2)(x_3\otimes y_3)\right)$ using \mref{prod_tens_coloré} lead to the following proposition.

\begin{proposition}\label{colour_cond}\cite{JZ} Let $\chi :\mathcal D\times\mathcal D\longrightarrow K$. The following are
equivalent\\
i) For $\mathcal A$, $\mathcal B$ $\mathcal D$-graded
    associative algebras, the product defined by
    (\ref{prod_tens_coloré}) is associative.\\
ii) $(\forall \alpha_1,\alpha_2,\alpha_3,\beta_1,\beta_2,\beta_3)\in\mathcal D$
\begin{equation}\label{cocycle}
      \chi (\beta_1,\alpha_2)\chi (\beta_1+\beta_2 ,\alpha_3) = \chi (\beta_2,\alpha_3)\chi (\beta_1,\alpha_2+\alpha_3)
\end{equation}
\end{proposition}

\begin{definition} Every mapping $\chi :\mathcal D\times\mathcal D\longrightarrow K$ which fulfills the equivalent conditions of proposition \mref{colour_cond} will be called a colour (twisting) factor.
\end{definition}

\begin{remarks} i) If $\chi$ is bilinear, which means in this context that the following equations are satisfied (for all $\al,\al',\be,\be'\in\mathcal{D}$)
\begin{eqnarray}\label{bichar}
      \chi (\alpha+\alpha ',\beta) &=& \chi (\alpha ,\beta)\chi (\alpha' ,\beta) \cr
      \chi (\alpha ,\beta+\beta ') &=& \chi (\alpha ,\beta)\chi
      (\alpha ,\beta'),
\end{eqnarray}
then the two members of (\ref{prod_tens_coloré}) amount to
    \begin{eqnarray}
      \chi (\beta_1, \alpha_2)\chi (\beta_1,\alpha_3)\chi (\beta_2,\alpha_3) &=& \prod_{1\leq i< j\leq 3} \chi (\beta_i,\alpha_j)
    \end{eqnarray}
and hence $\chi$ is a colour factor \footnote{These bilinear mappings are also called bicharacters in the literature \cite{Ree}.}. But the full class of colour factors is much larger than solutions of Eq. \mref{bichar}. Just observe that Eq. \mref{cocycle} is 
homogeneous in the classical sense {\em i.e.} for all $\lambda\in K$, if $\chi$ fulfills \mref{cocycle} then rescaling it by $\lambda$ still does. Hence, for example, any constant function on $\mathcal{D}\times \mathcal{D}$ is a colour factor. This shows the existence of colour factors that are not bilinear.\\
ii) The converse (i.e., $ii\Longrightarrow i$) part of proposition \mref{colour_cond} can be easily proved by considering (free) semigroup algebras $K[\mathcal{D}]$.
\end{remarks}

\begin{note}\label{ass_and_morph}
i) The colour product of two algebras $\displaystyle{\mathcal A={\oplus}_{\alpha\in \mathcal D}\mathcal A_\alpha}$ and 
$\displaystyle{\mathcal B=\oplus_{\beta\in \mathcal D}\mathcal B_\beta}$ comes also as a graded algebra by 
\begin{equation}\label{gr_of_prod}
	(\mathcal A\otimes \mathcal B)_\ga=\oplus_{\al+\be=\ga}\ \mathcal A_\al\otimes \mathcal B_\be .
\end{equation}

The usual identification 

\begin{equation}
	(\mathcal A\otimes \mathcal B)\otimes \mathcal C\simeq \mathcal A\otimes (\mathcal B\otimes \mathcal C)
\end{equation}
holds for coloured products.\\
ii) Moreover, if $\mathcal A \stackrel{f}{\ra}\mathcal A'$ (resp. $\mathcal B \stackrel{g}{\ra}\mathcal B'$) are two morphisms of (graded) algebras (over the same semigroup of degrees $\mathcal{D}$), then 
$\mathcal A\otimes \mathcal B \stackrel{f\otimes g}{\longrightarrow}\mathcal A'\otimes \mathcal B'$ is a morphism of algebras (the colour products being taken w. r. t. the same colour factor).
\end{note}

\section{Special classes of laws}
\subsection{Dual laws}
\subsubsection{Algebras and coalgebras in duality}\label{dual_laws}

An algebra $(\A,\mu)$ and a coalgebra $(C,\Delta)$ are said to be in duality iff there is a non-degenerate pairing $\scal{-}{-}$ such that for all $x,y\in\A,\ z\in C$
\begin{equation}
	\scal{\mu(x,y)}{z}=\scal{x\otimes y}{\Delta(z)}^{\otimes 2}
\end{equation}
In the following, we will call \textit{dual law} a product $\ncp{K}{A}\otimes \ncp{K}{A}\stackrel{*}{\longrightarrow} \ncp{K}{A}$ on the free algebra which is the dual of a comultiplication, the pairing being given on the basis of words by $\scal{u}{v}=\delta_{u,v}$.

\bs
Our first examples are essential in modern and not-so-modern research (\cite{Oc,Ro}). Firstly, we have the dual of the Cauchy product
\begin{equation}\label{coCauchy}
	\Delta_{Cauchy}(w)=\sum_{uv=w} u\otimes v\ .
\end{equation}
Contrary to this one \mref{coCauchy}, which is not a morphism of algebras\footnote{Unless $A=\emptyset$.} 

\begin{equation}\label{free_alg}
\ncp{K}{A}\longrightarrow \ncp{K}{A}\otimes \ncp{K}{A}\ ,
\end{equation}

one has three very well-known examples being so, namely duals of the shuffle $\shuffle$, the Hadamard $\odot$ and the infiltration product $\ua$. As they are morphisms between the algebras \mref{free_alg}, they are well defined by their values on the letters. Respectively 
\begin{equation}
\Delta_\shuffle(x)=x\otimes 1+ 1\otimes x\ ;\ \Delta_\odot(x)=x\otimes x\ ;\ \Delta_\ua(x)=x\otimes 1+ 1\otimes x + x\otimes x\ .
\end{equation}
One can prove that the deformations $\Delta_q=\Delta_\shuffle(x)+ q \Delta_\odot(x)$ are also co-associative and that they are the unique solutions of the problem of bialgebra comultiplications on $\ncp{K}{A}$ that are compatible with subalphabets \cite{DFLL}.

In the sequel, we will make use several times of the following lemma, the proof of which is left to the reader.

\begin{lemma}\label{alg_coalg} Let $\mathcal{A}$ be an algebra and $\mathcal{C}$ be a coalgebra in (non-degenerate) duality, then $\mathcal{A}$ is associative iff $\mathcal{C}$ is coassociative.
\end{lemma}

\subsubsection{Duality between grouplike elements and unities}\label{dual_unit_grouplike}

Let $(\mathcal{C},\Delta)$ be a coalgebra with counit $\ep : \mathcal{C}\ra K$. We call \textit{group-like} an element $u$ such that 
\begin{equation}
	\ep(u)=1\ ;\ \Delta(u)=u\otimes u\ .
\end{equation}

One then has $\mathcal{C}=ker(\ep)\oplus K.u$ and 
\begin{equation}\label{delta+}
\Delta(y)=\Delta^+(y)+y\otimes u+u\otimes y-\ep(y)u\otimes u\ .
\end{equation}
where $\Delta^+$ is a comultiplication on $\mathcal{C}$ for which $ker(\ep)=\mathcal{C}^+$ is a subcoalgebra (i.e., $\Delta^+(\mathcal{C}^+)\subset \mathcal{C}^+\otimes \mathcal{C}^+$) \cite{B_Lie_I_III}.

\begin{proposition} Let $(\mathcal{C},\Delta,\ep)$ be a coalgebra with counit, $u$ a group-like element in $\mathcal{C}$ and $(\mathcal{C}^+,\Delta^+)$ be as in \mref{delta+}. On the other hand, let $\A$ be an algebra and $\A^{(1)}=\A\oplus K.v$ be the algebra with unit constructed from $\A$ by adjunction of the unity $v$. Then, if $\mathcal{C}^+$ and $\A$ are in duality by $\scal{\ }{\ }$, so are $\mathcal{C}$ and $\A^{(1)}$ by $\scal{\ }{\ }_\bu$ defined as follows
\begin{equation}
	\scal{x+\al v}{y+\be u}_\bu=\scal{x}{y}+\be\al
\end{equation}
for $x\in \A$ and $y\in \mathcal{C}^+=ker(\ep)$.
\end{proposition}
\Proof Let 
\begin{eqnarray}\label{eq14}
	\scal{(x_1+\al_1 v)\otimes (x_2+\al_2 v)}{\Delta(y+\be u)}_\bu^{\otimes 2}=\cr
	\scal{(x_1+\al_1 v)\otimes (x_2+\al_2 v)}{\Delta^+(y)+ y\otimes u+u\otimes y+\be u\otimes u)}_\bu^{\otimes 2}
\end{eqnarray}
but, according to the fact that 
$$
\scal{x_i}{u}=\scal{x_1\otimes v}{\Delta^+(y)}=\scal{v\otimes x_2}{\Delta^+(y)}=\scal{v\otimes v}{\Delta^+(y)}=\scal{v}{y}=0
$$
 one has from \mref{eq14}
\begin{eqnarray}
&&\scal{(x_1+\al_1 v)\otimes (x_2+\al_2 v)}{\Delta(y+\be u)}_\bu^{\otimes 2}=\cr
	&&\scal{x_1\otimes x_2}{\Delta^+(y)}^{\otimes 2} +\al_2 \scal{x_1}{y}+ 
	\al_1 \scal{x_2}{y}+\al_1\al_2\be=\cr
	&&\scal{x_1x_2+\al_2 x_1+\al_1 x_2+\al_1\al_2 v}{y+\be u}_\bu=\scal{(x_1+\al_1 v)(x_2+\al_2 v)}{y+\be u}_\bu
\end{eqnarray}
which proves the claim.\cqfd

\subsection{Deformed laws}\label{Deformed laws}

Let $S$ be a semigroup graded by a semigroup of degrees $\mathcal{D}$ and $\A=K[S]$ its algebra. A colour factor 
$\chi : \mathcal{D}\times \mathcal{D}\ra K$ being given, we endow the algebra $\A\otimes \A$ with the coloured tensor product structure. Notice that the diagonal subspace $D_S=\oplus_{x\in S}K x\otimes x$ is a subalgebra as 
\begin{equation}\label{eq13}
(x\otimes x)(y\otimes y)=\chi(x,y)\ xy\otimes xy\ .
\end{equation}
Carrying \mref{eq13} back to $\A$ by means of the isomorphism of vector spaces, $\A\ra D_S$, one sees immediately that the deformed product on $\A$ given by 
\begin{equation}\label{def_prod}
x._\chi y=\chi(x,y)\ xy
\end{equation}
(for $x,y\in \mathcal{D}$) is associative.\\
From now on, we suppose that the semigroup $\mathcal{D}$ fulfils condition [D] of Bourbaki \cite{B_Alg_I_III}
which means that for all $z\in \mathcal{D}$, the number of solutions $(x,y)\in\mathcal{D}^2$ of the equation 
$xy=z$ is finite. This condition is fulfilled by almost all the grading semigroups used by combinatorialists, 
in particular the semigroups $(\N,+),\ (\N^+,\times), (\N^{(X)},+)$.

\ss
If $\A$ is endowed with the scalar product for which the basis 
$(s)_{s\in S}$ is orthonormal, the pairing is non-degenerate and the dual comultiplication is given by 
\begin{equation}
\Delta(z)=\sum_{xy=z}\chi(x,y)\ x\otimes y\ .
\end{equation}
The construction together with lemma \ref{alg_coalg} proves that this comultiplication on $\A$ is coassociative.

\subsection{Shifted laws}

We begin by a very general version of the ``shifting lemma'' (more general than the one given and needed in \cite{GoF12}).\\ 
We start from an algebra $\A$ decomposed (as a vector space) by the direct sum 
$$
\A=\oplus_{\al\in \D} \A_\al
$$ 
over $\D$, a semigroup. We denote by $\mathrm{End}^{\al}(\A)$ the morphisms of algebra 
$\A\ra\A$ (then multiplicative) which ``shift by $\alpha$'' (i.e., $\phi\in \mathrm{End}^{\al}(\A)$ 
iff for all $\be\in \mathcal{D}$, one has\\ 
$\phi(\A_\be)\subset \A_{\al+\be}$). this situation is typical of ``shift of indices'' in free algebras. 

\ss
For example let $Y=\{y_j\}_{j\geq 1}=\{y_1,y_2,\cdots y_k,\cdots \}$ be an infinite alphabet, $\mathcal{D}$ be its 
indexing semigroup $(\N^+,+)$ and $\A$ be the algebra $\ncp{K}{Y}$. For every monomial (word) $w$ let $d(w)$ be the 
maximal index $j$ of a letter $y_j$ occurring in $w$. With $\ncp{K}{Y}_j:=\oplus_{d(w)=j}K.w$, one gets a direct sum decomposition 
\begin{equation}
	\ncp{K}{Y}=\oplus_{j\geq 1}\ncp{K}{Y}_j
\end{equation}

for which $\ncp{K}{Y}$ is not a graded algebra (it is, in fact for the $sup$ law in $\N^+$, but we aim here at constructing a graded algebra for the addition of degrees). The change of variables $T_n(y_j):=y_{j+n}$ defines a morphism of algebras $T_n\in \mathrm{End}^{n}(\A)$. the product of algebra is the usual concatenation whereas the shifted law reads 
$$
w_1\ \overline{conc}\ w_2:=w_1T_n(w_2)
$$
where $n=max\{j\geq 1||w_1|_{y_j}\not=0\}$.\\
One can easily check that the following spaces are subalgebras of $(\ncp{K}{Y},\overline{conc})$
\begin{enumerate}
	\item the space generated by packed words (i.e., the words whose alphabet indices are of the form $[1\ldots q]$)
	\item the space generated by injective words (each letter occurs at most once)
	\item the space generated by permutation words (packed and injective, see \cite{ncsf3})
	\item the space generated by increasing (resp. strictly increasing) words (i.e., $w=y_{j_1}y_{j_2}\cdots y_{j_k}$ such that the function $r\ra j_r$ is increasing (resp. strictly increasing)
	\item the space generated by disconnected words (i.e., $w=y_{j_1}y_{j_2}\cdots y_{j_k}$ such that there exists en index $r<k$ with $y_{j_r+1}$ not occurring in $w$)
\end{enumerate}
The following lemma gives general conditions for such shifted laws to be associative. 

\begin{lemma}
Let $\A$ be an algebra (whose multiplicative law will be denoted by $\star$) and $\A=\oplus_{\al\in \D} \A_\al$ 
a decomposition of $\A$ as a direct sum over $\D$, a semigroup 
($\A$ is then graded but only as a vector space).
Let $\al\mapsto T_\al$: $\D \ra \mathrm{End}^{gr}(\A)$ be a morphism of semigroups such that $T_\al\in \mathrm{End}^{\al}(\A)$. 
Explicitly, for all $\al,\be\in \D\ ;\ x\in\A_\be$  
\begin{equation}
	T_\al(x)\in \A_{\al+\be} \textrm{ and } T_\al\circ T_\be=T_{\al+\be}\ .
\end{equation}
We suppose that the shifted law defined for $x\in \A_\al$ and $y\in \A$ by
\begin{equation}\label{shifted}
    x\ \bar\star\ y=x\star T_\al(y)
\end{equation}
is graded for the decomposition 
$$
\A=\oplus_{\al\in \D} \A_\al
$$ 
(i.e., if $x\in \A_\al$ and $y\in \A_\be$ then $x\ \bar\star\ y\in \A_{\al +\be}$).\\
Then, if the law $\star$ is associative so is the law $\bar\star$.
\end{lemma}

\Proof One has just to prove the identity of associativity of $\bar\star$ for homogeneous elements. 
Suppose that $\star$ is associative, for $x\in \A_\al,\ y\in \A_\be$ and $z\in \A$, one has
\begin{eqnarray*}
x\bar\star(y\bar\star z)=x\star T_\al(y\bar\star z)=x\star T_\al(y\star T_\be (z)))=x\star (T_\al(y)\star T_\al(T_\be (z)))=\cr
x\star (T_\al(y)\star T_{\al+\be} (z))=
(x\star T_\al(y))\star T_{\al+\be}(z)=
\underbrace{(x\ \bar\star\ y)}_{\in \A_{\al+\be}}\star\ T_{\al+\be}(z)=(x\bar\star y)\bar\star z
\end{eqnarray*}
\cqfd

\section{Application to the structure of $\LDIAG(q_c,q_s)$}

\subsection{Associativity of $\LDIAG(q_c,q_s)$ using the previous tools}

As was stated in paragraph \ref{code_to_rec}, we just have to prove proposition \ref{deformed_law} and we keep the notations of it. We first remark, from paragraph \mref{Deformed laws} that, for a semigroup $S$ of type (D)\footnote{After \cite{B_Alg_I_III}, a semigroup $S$ of type (D) is such that the product mapping $S\times S\ra S$ has finite fibers.}, graded by a degree function $|\ |_d:\ S\ra \N$, the comultiplication $\Delta_1:\ K[S]\ra K[S]\otimes K[S]$
given for $s\in S$ by
\begin{equation}
	\Delta_1(s)= \sum_{rt=s} q_s^{|r|_d|t|_d} r\otimes t
\end{equation}
is coassociative.\\
Now, we endow $\ncp{K}{S}\otimes \ncp{K}{S}$ with the structure of coloured product given by the bicharacter on $S^*$ 
\begin{equation}
	\chi(u,v)=\prod_{1\leq i \leq |u|\atop 1\leq j \leq |v|} q_c^{|u[i]|_d|v[j]|_d}
\end{equation}
One defines a mapping $\Delta:\ S\ra \ncp{K}{S}\otimes \ncp{K}{S}$ by
\begin{equation}
	\Delta(s)=s\otimes 1_{S^*}+1_{S^*}\otimes s+\Delta_1(s)
\end{equation}
which is extended at once as a morphism of algebras $\Delta:\ \ncp{K}{S}\ra \ncp{K}{S}\otimes \ncp{K}{S}$. Note that $V=\oplus_{x\in S\cup \{1_{S^*}\}}K.x=KS\oplus K.1_{S^*}$ is a subcoalgebra for $\Delta$ and the coalgebra $V$ is, by paragraph \mref{dual_unit_grouplike}, still coassociative. 
Now, one has to prove that the following rectangle is commutative

\begin{eqnarray}\label{co-associativity}
\xymatrix{
\ar[rr]^{\Delta}\ncp{K}{S}\ar[d]_{\Delta}&&\ar[d]^{Id\otimes \Delta}\ncp{K}{S}\otimes\ncp{K}{S}\\
 \ncp{K}{S}\otimes\ncp{K}{S}\ar[rr]^{\Delta\otimes Id}&&\ncp{K}{S}\otimes\ncp{K}{S}\otimes\ncp{K}{S}
}\ .
\end{eqnarray}

By Note \mref{ass_and_morph} (ii) all the arrows are morphisms of algebras and in particular the composites $(Id\otimes \Delta)\circ \Delta\ ;\ (\Delta\otimes Id)\circ \Delta$ which, it has been proved just previously, coincide on $S$ (coassociativity of the subcoalgebra $V$). This shows that the rectangle \mref{co-associativity} is commutative.\\

\textbf{End of the duality}\pointir\\
We denote by $\da$ the law which is dual to $\Delta$. This law, being dual to a coassociative comultiplication, is associative. We prove that it satisfies the same recursion as in proposition \mref{deformed_law} so, $\da=\ua$. It is sufficient to prove the recursion for non-empty factors. One has

\begin{eqnarray}
au \da bv	=\sum_{w\in S^+}\scal{au \da bv	}{w} w=\sum_{x\in S\atop w_1\in S^*}\scal{au \da bv	}{xw_1} xw_1=
\sum_{x\in S\atop w_1\in S^*}\scal{au \otimes bv	}{\Delta(x)\Delta(w_1)} xw_1=\cr
\sum_{x\in S\atop w_1\in S^*}\scal{au \otimes bv	}{(x\otimes 1+1\otimes x+\sum_{yz=x}\chi(y,z) y\otimes z)\Delta(w_1)} xw_1=\cr
\sum_{x\in S\atop w_1\in S^*}\scal{au \otimes bv	}{(x\otimes 1)\Delta(w_1)} xw_1 +
\sum_{x\in S\atop w_1\in S^*}\scal{au \otimes bv	}{(1\otimes x)\Delta(w_1)} xw_1 +\cr
\sum_{x\in S\atop w_1\in S^*}\scal{au \otimes bv	}{(\sum_{yz=x}\chi(y,z) y\otimes z)\Delta(w_1)} xw_1=\cr
\sum_{x=a\atop w_1\in S^*}\scal{au \otimes bv	}{(x\otimes 1)\Delta(w_1)} xw_1 +
\sum_{x\in S\atop w_1\in S^*}\scal{au \otimes bv	}{(1\otimes x)\sum_{i,j}\be_{ij}w_i\otimes w_j} xw_1 +\cr
\sum_{x\in S\atop w_1\in S^*}\scal{au \otimes bv	}{(\sum_{yz=x}\chi(y,z) y\otimes z)\sum_{i,j}\be_{ij}w_i\otimes w_j} xw_1=\cr
a(u\da bv)+ q_c^{|au|_d|b|_d} b(au\da v)+ 
q_c^{|u|_d|b|_d}q_s^{|a|_d|b|_d} (a . b)(u\da v)\ \ 
\end{eqnarray}

which proves the claim.

\subsection{Structure of $\LDIAG(q_c,q_s)$}

This section is devoted to the thorough study of the structure of $\LDIAG(q_c,q_s)$ using that of the algebra of $(\mathfrak{MON}^+(X))^*$ endowed with the shifted law $\bar\ua$.\\ 
We first investigate the structure of the monoid $((\mathfrak{MON}^+(X))^*,\bar\star)$, broadening out to some extent Proposition 3.1 of \cite{GoF12}. 
For a general monoid, $(M,\star,1_M)$, the irreducible elements are the elements $x\neq 1_M$ such that $x=y\star z\Longrightarrow 1_M\in \{y,z\}$. The set of these elements will be denoted $irr(M)$. For convenience, in the following statement, $M$ stands for the monoid $((\mathfrak{MON}^+(X))^*,\bar\star)$, $M^+=M-\{1_M\}$ and $M_c$ is the submonoid of codes of diagrams (i.e., words which fulfill Eq. \mref{criterium_code_of_a_diagram}).\\
The monoid $M$ is free. An element $w=m_1.m_2.\cdots .m_l\in M^+$ (hence $l=|w|>0$) is reducible iff there exists $0<k<l$ such that 
\begin{equation}
\Big(indices(Alph(m_1.m_2.\cdots .m_k)))\prec (indices(Alph(m_{k+1}.m_{k+2}.\cdots .m_l))\Big)
\end{equation}
where, for two nonempty subsets $X,Y\subset \N^+$, one notes $\prec$ the relation of majoration i.e., 
\begin{equation}
	(\forall (x,y)\in X\times Y)(x<y)\ .
\end{equation}
One checks at once that the monoid $M_c$ is generated by the subalphabet $irr(M)\cap M_c$ and therefore is free.\\
Now, we need a classical tool of general algebra (see \cite{B_Comm_Alg_I_III}, chapter III for details).\\
Let $(\A,\mu)$ be an algebra endowed with an increasing exhaustive filtration $(\A_n)_{n\in \N}$ (i.e., two-sided ideals such that $\A_n\subset \A_{n+1}$ and $\cup_{n\in \N} \A_n=\A$). It is classical to construct the associated graded algebra $Gr(\A)=\oplus_{n\geq 0}\A_n/\A_{n-1}$ by passing the law to quotients i.e., $\bar\mu_{p,q}\ :\ \A_p/\A_{p-1}\otimes \A_q/\A_{q-1}\ra \A_{p+q}/\A_{p+q-1}$ (one sets $\A_{-1}=\{0\})$. A classical lemma (and easy exercise) states that, if the associated graded algebra is free, so is $\A$.\\
Now, returning to $(\ncp{K}{\mathfrak{MON}^+(X)},\bar\ua)$ ($\bar\ua$ is the shifted deformed law), one constructs a filtration by the number of irreducible components of a word of monomials (call it $l(w)$ for $w\in \mathfrak{MON}^+(X)$). From \mref{deformed_infiltr}, one gets, 
\begin{equation}
	w_1\bar\ua w_2=w_1\bar\star w_2+ \sum_{l(w)<l(w_1)+l(w_2)} P_w(q_c,q_s) w 
\end{equation}
with $P_w\in K[q_c,q_s]$ (indeed, $\bar\ua$ is the same law as in \mref{deformed_infiltr} but shifted) and then, the associated graded algebra is, by triangularity argument, free. One can then state the following structure theorem.

\begin{theorem}
The algebra $\ncp{K}{\mathfrak{MON}^+(X)}$, endowed with the shifted deformed law $\bar\ua$, is free on the irreducible words and then the algebra $\LDIAG(q_c,q_s)$, isomorphic to a subalgebra generated by irreducible words, is free for every choice of $(q_c,q_s)$.
\end{theorem}

\section{Conclusion}

To sum up what has been done in this paper we can state that the deformed algebra $\LDIAG(q_c,q_s)$, which originates from a special quantum field theory \cite{BBM}, is free and its law can be constructed from very general procedures: it is a shifted twisted law. Before shifting, one can observe that the law is, in fact, dual to a comultiplication on a free algebra. This comultiplication is a perturbation, with $q_s$ (the superposition parameter) of the shuffle comultiplication on this free algebra. The parameter $q_s$ is obtained by addition of a perturbating factor which is just dual to a (diagonally) deformed law of a semigroup whereas the crossing parameter $q_c$ is obtained by extending to the tensor structure (i.e., to words) a colour factor of an algebra.

\end{document}

%% file: DTPK_fig1.tex
\ifx\JPicScale\undefined\def\JPicScale{1}\fi
\unitlength \JPicScale mm

\begin{picture}(122.5,47.5)(0,46)
\linethickness{0.75mm}
\multiput(69.34,49.61)(0.12,0.37){76}{\line(0,1){0.37}}
\linethickness{0.75mm}
\multiput(72.23,52.11)(0.12,0.37){67}{\line(0,1){0.37}}
\linethickness{0.75mm}
\multiput(82.76,78.82)(0.12,-0.57){48}{\line(0,-1){0.57}}

\put(60,80){\circle{5}}
\put(80.13,79.87){\circle{5}}
\put(100.13,79.87){\circle{5}}
\put(120,80){\circle{5}}

\put(70.2,49.88){\circle*{5}}
\put(90.3,49.56){\circle*{5}}
\put(110,50){\circle*{5}}

\put(59,86){$1$}
\put(79,86){$2$}
\put(99,86){$3$}
\put(119,86){$4$}

\put(69,41.88){$1$}
\put(89.3,41.88){$2$}
\put(109,41.88){$3$}

\linethickness{0.75mm}
\multiput(71.97,49.74)(0.12,0.14){212}{\line(0,1){0.14}}

\linethickness{0.75mm}
\multiput(102.76,79.47)(0.12,-0.48){57}{\line(0,-1){0.48}}

\linethickness{0.75mm}
\multiput(92.76,49.61)(0.12,0.36){81}{\line(0,1){0.36}}
\linethickness{0.75mm}
\multiput(90.26,48.29)(0.12,0.37){78}{\line(0,1){0.37}}
\linethickness{0.75mm}
\multiput(88.29,49.08)(0.12,0.36){81}{\line(0,1){0.36}}
\linethickness{0.75mm}
\multiput(111.97,50.66)(0.12,0.37){72}{\line(0,1){0.37}}
\linethickness{0.75mm}
\multiput(109.74,51.05)(0.12,0.37){73}{\line(0,1){0.37}}
\linethickness{0.75mm}
\multiput(61.71,78.03)(0.12,-0.13){216}{\line(0,-1){0.13}}
\end{picture}

\vspace{0.8cm}
\begin{center}
{\small{\bf Fig 1}\pointir \it Labelled diagram of format $3\times 4$.} 
\end{center}
%

%% file: DTPK_fig2.tex
\ifx\JPicScale\undefined\def\JPicScale{1}\fi
\unitlength \JPicScale mm

\begin{center}
\begin{tabular}{|c||c|c|c|c|c|c|c|}
\hline
$j$ 				& $2$ & $3$ & $1$ & $2$ & $3$ & $3$ & $4$\\ 
\hline
$i$ 				& $1$ & $1$ & $2$ & $2$ & $2$ & $3$ & $3$\\
\hline
$\om(i,j)$  & $2$ & $1$ & $1$ & $1$ & $3$ & $1$ & $2$\\
\hline
\end{tabular}
\end{center}

\begin{center}
{\small{\bf Fig 2}\pointir \it The weight function (when not $0$) of the diagram in Fig 1. Here $p=3$ and $q=4$.} 
\end{center}